# Contrast Enhancement of Brightness-Distorted Images by Improved Adaptive Gamma Correction

**Gang Cao[1*], Lihui Huang[1], Huawei Tian[2], Xianglin Huang[3], Yongbin Wang[1], Ruicong Zhi[4]**

[1]School of Computer Science, Communication University of China, Beijing 100024, China
[2]College of Criminal Investigation and Counter Terrorism, People's Public Security University of China, Beijing 100038, China
[3]Faculty of Science and Technology, Communication University of China, Beijing 100024, China
[4]School of Computer and Communication Engineering, Beijing University of Science and Technology, Beijing 100083, China
[*]Correspondence author: gangcao@cuc.edu.cn

**Abstract.** As an efficient image contrast enhancement (CE) tool, adaptive gamma correction (AGC) was previously proposed by relating gamma parameter with cumulative distribution function (CDF) of the pixel gray levels within an image. ACG deals well with most dimmed images, but fails for globally bright images and the dimmed images with local bright regions. Such two categories of brightness-distorted images are universal in real scenarios, such as improper exposure and white object regions. In order to attenuate such deficiencies, here we propose an improved AGC algorithm. The novel strategy of negative images is used to realize CE of the bright images, and the gamma correction modulated by truncated CDF is employed to enhance the dimmed ones. As such, local over-enhancement and structure distortion can be alleviated. Both qualitative and quantitative experimental results show that our proposed method yields consistently good CE results.

## 1. Introduction

Contrast enhancement (CE) refers to the image enhancement on contrast by adjusting the dynamic range of pixel intensity distribution [1]. CE plays an important role in the improvement of visual quality for computer vision, pattern recognition and digital image processing. In real applications, we usually encounter digital images with poor contrast or abnormal brightness, which may result from different factors, such as the inexperience of taking photographs and the inherent deficiency of imaging devices. The capturing scenes with low or high illuminance intensity may also lead to reduced contrast quality. Despite of visual quality degradation, low contrast might hinder the further applications of a digital image, including image analysis and understanding, object recognition and digital printing, etc. As such, it is essential to enhance the contrast of such distorted images before further applications.

Existing CE techniques can be categorized into pixel-domain [2-12] and transform-domain ones [13-17] according to the data domain they are applied to [6]. The former relies on pixel intensity operation, while the latter implements CE in the transformation domain of an image, such as discrete cosine transform (DCT) [13-15], Wavelet [16] and Curvelet [17]. Generally, the pixel-domain CE techniques might be used more widely in real applications due to low requirements on computational cost and parameter setting.

There exist a large category of pixel-domain CE techniques based on the redistribution of gray levels, such as histogram equalization (HE) and its related methods [1-3]. HE implements pixel intensity mapping by directly equalizing the cumulative distribution function (CDF) of the input image's gray level histogram, which becomes as uniform as possible after CE. Despite the merit of high computational efficiency, HE owns the limitation to incur over-enhancement if high peaks exist in the input histogram [4]. In order to attenuate such deficiency, the improved local HE [2] and brightness preserving bi-HE [3] are developed. As another influential work, Arici *et al.* propose a general histogram modification (HM) framework for CE, which is considered as an optimization problem [4]. It minimizes a cost function which includes the penalty of histogram deviation from primary to uniform histograms, histogram smoothness and black & white stretching. Although HM successfully avoids the unnatural look caused by excessive enhancement, its enhancement results are rather sensitive to parameter setting. Gaussian mixture model is also proposed to model the image intensity distribution which is partitioned into several intervals [5]. Pixel gray levels in each interval are mapped to the appropriate output interval according to dominant Gaussian component and interval-wise CDF. Although good visual effects are gained, such a method has a high computational cost.

Recently, the spatial entropy-based contrast enhancement (SECE) is proposed to incorporate the spatial distribution of pixel intensities into the design of mapping function [6]. After dividing an input image into non-overlapped blocks, the distribution of spatial entropy is first calculated from blockwise 2D spatial histograms, and then equalized for implementing CE. SECE can consistently yield visually improved and pleasing outputs without attractive distortions, regardless of the available contrast on input images. However, as pointed out in the later improvement works [7, 8], SECE is found to enjoy the insufficient enhancement strength limit. In [9], noisy low-light images are enhanced by recurring to structure-texture-noise decomposition model of images. Besides, human visual perception [18, 19] and advanced machine learning techniques, such as deep learning [20], are also used to design efficient CE algorithms.

Gamma correction is also a popular pixel-domain CE method, which is cost-effective and good at dealing with bright and dimmed images [1, 11, 12, 21]. However, the manual selection of appropriate gamma values is often time-consuming. As for adaptive gamma correction (AGC), the gamma parameter is modulated by the statistics extracted from images, and therefore set automatically. Huang *et al.* proposed the AGC with weighting distribution (AGCWD) by setting gamma as a function of CDF [11]. AGCWD behaves well in enhancing dimmed images which own low average brightness and seem black, as shown in Fig. 1(a).

In this paper, we focus on the CE of brightness-distorted images which own a relatively high or low global intensity. The existing AGC techniques are revisited and improved formally. We find that such methods are incapable to be directly used to enhance globally bright images, and the image structure in local bright regions may be lost in enhancing dimmed images. In order to attenuate such deficiencies, we propose an improved AGC method by integrating the strategies of negative images and CDF truncation. Substantial test results verify the effectiveness and efficacy of our proposed method in enhancing both dimmed and bright types of contrast-distorted images.

The rest of this paper is organized as follows. AGC is revisited and analyzed detailedly in Section 2, followed by the improved AGC scheme proposed in Section 3. Section 4 shows experimental results. The conclusion is drawn in Section 5.

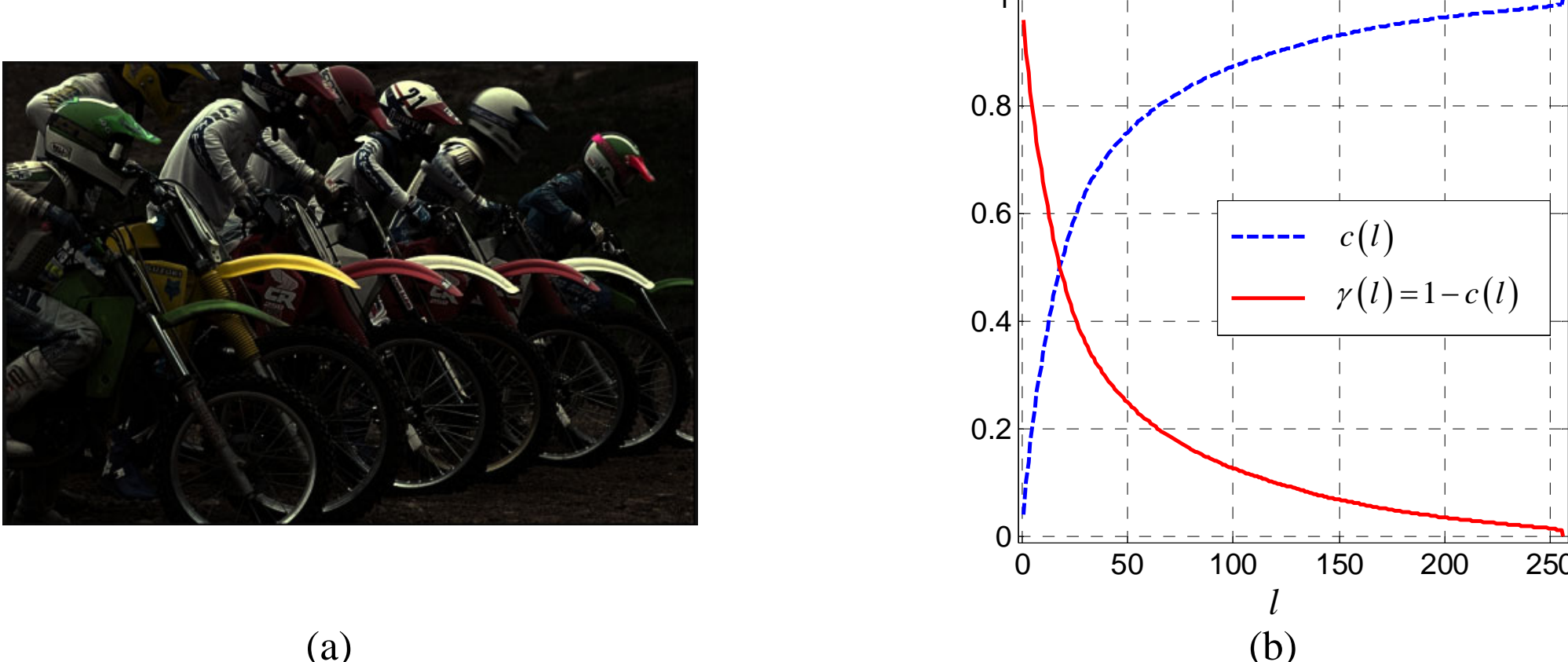

(a) (b)

**Fig. 1** CDF-based adaptive gamma parameter. (a)(b) Example dimmed image and its adaptive gamma curve. Here, the intensity channel is referred.

## 2. Prior Works on Adaptive Gamma Correction

In [11], a typical AGC method is proposed by relating gamma parameter with CDF. The transformed pixel intensity T($l$) is computed as

$$\mathrm{T}(l) = round\left[ l_{\max} \left( \frac{l}{l_{\max}} \right)^{\gamma(l)} \right]. \tag{1}$$

where $\gamma(l)$=1-$c(l)$=1-$\sum_{x=0}^{l} p(x)$, $l$=0, 1, 2, ..., 255 is the CDF of gray levels in the input image. $p(x)$ denotes the normalized gray level histogram. $round[\cdot]$ is the rounding operation. Here, 8-bit grayscale images with maximum pixel intensity $l_{\max}$=255 are exampled. As shown in Fig. 1(b), the gamma value $\gamma(l)$ monotonically decreases from 1 to 0. As such, most of the low-valued pixels in dimmed images will be stretched. The pixel dynamic range of resulted images will be extended, so that CE can be achieved.

Furthermore, a weighting distribution function is used to smooth the primary histogram [11]. That is,

$$p_w(l) = p_{\max}\left(\frac{p(l) - p_{\min}}{p_{\max} - p_{\min}}\right)^{\alpha} \tag{2}$$

where $\alpha$ is the adjusted parameter, $p_{\max} = \max_l p(l)$, $p_{\min} = \min_l p(l)$. Then $p_w(l)$ is normalized for yielding $p_w^{'}(l)$. As shown in Fig. 2(a), the histogram weighted by $\alpha < 1$ becomes relatively flat and smooth at low intensities, so that the adverse effects are lessened [11].

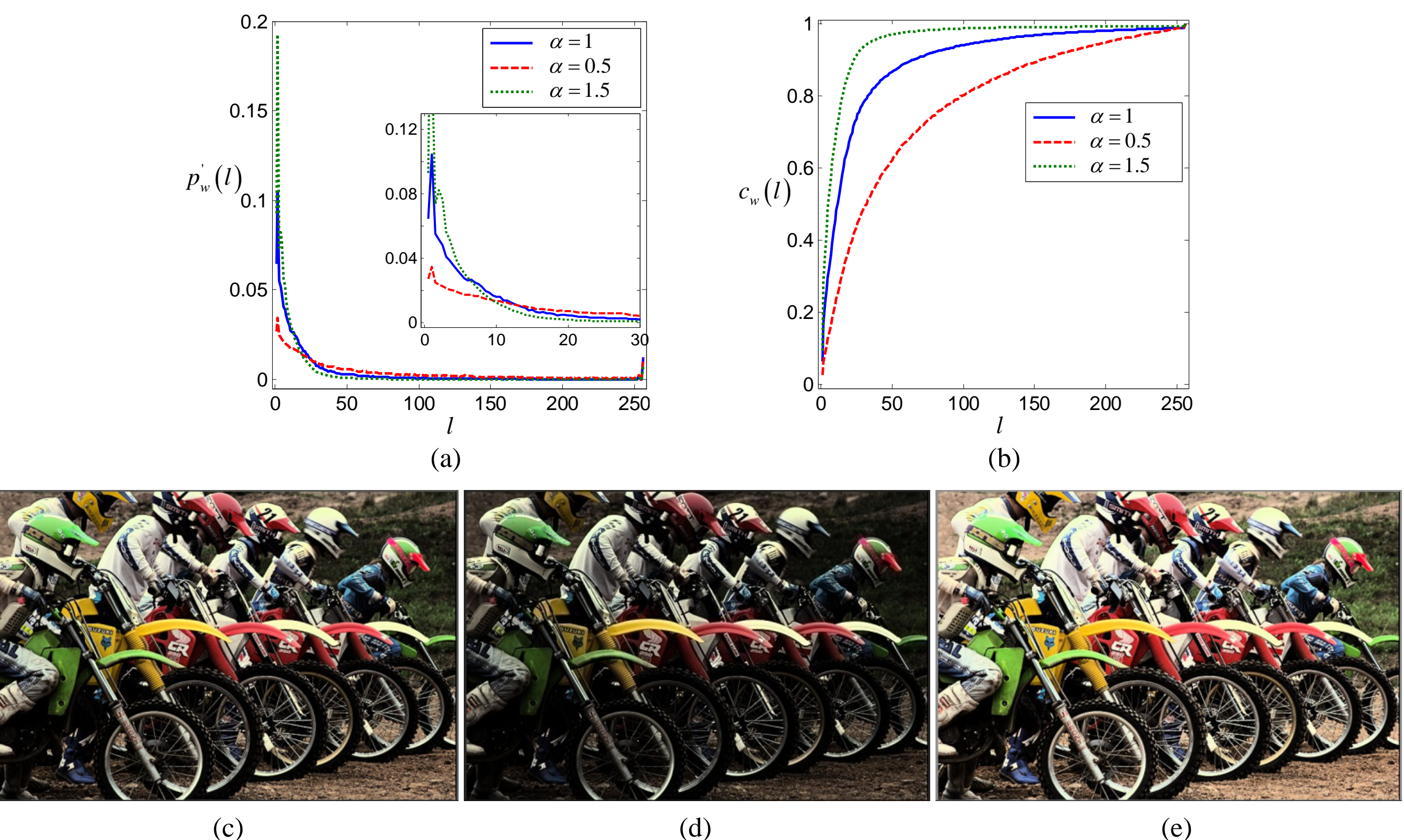

**Fig. 2** AGCWD method. (a) The weighted pixel intensity histogram of the image in Fig 1(a); (b) its corresponding CDF curve; (c)(d)(e) enhanced images with the weighting factor $\alpha$ = 1, 0.5, 1.5, respectively.

In the formal AGCWD algorithm [11], Eq. (1) is updated by replacing $\gamma(l)$ with the $\gamma_w(l)$ derived from $p_w^{'}(l)$. As shown in Fig. 2(b), the corresponding CDF $c_w(l)$ from $p_w^{'}(l)$ can be adjusted via $\alpha$. Figs. 2(c)(d)(e) show the corresponding enhancement results of AGCWD.

Although such prior AGC methods are effective in enhancing the contrast of most dimmed images, the image distortion may be incurred in bright regions due to the improper setting of rather low gamma values for large pixel intensities. Moreover, such AGC methods fail to enhance the contrast of globally bright images, which own distinctly different distribution of pixel intensities.

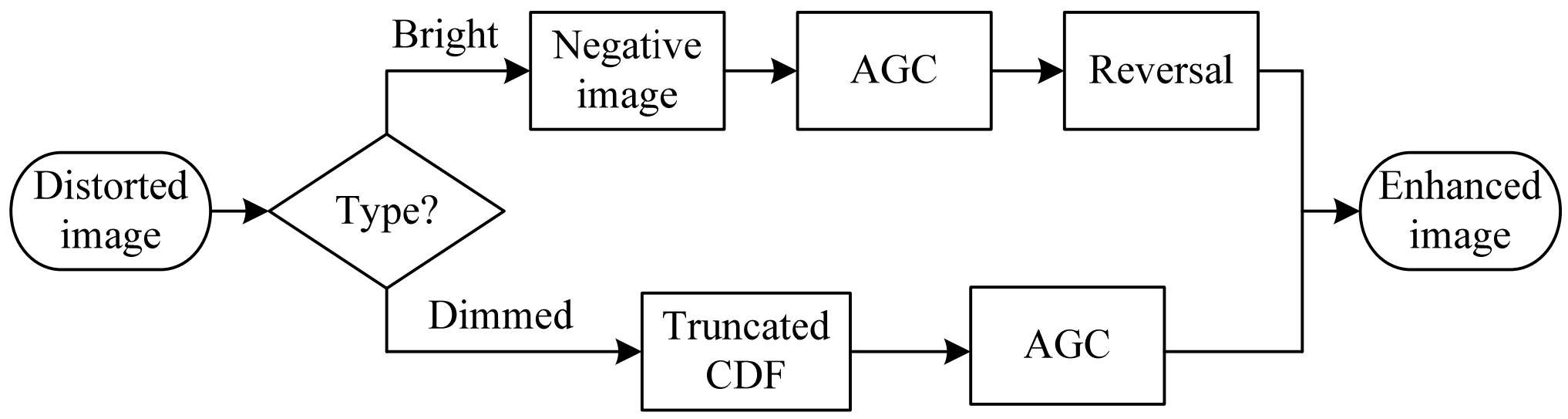

**Fig. 3** Flowchart of the proposed image contrast enhancement algorithm.

## 3. Proposed Method

The overview of our proposed CE scheme is indicated in Fig. 3. It should be pointed out that here we focus on the CE of two types of brightness-distorted images, i.e., the dimmed and the bright ones. As for an input image $\mathbf{I}(x, y)$, $x$=1, 2, ..., $M$, $y$=1, 2, ..., $N$, its type is first identified by thresholding the statistical quantity

$$t = \frac{m_{\mathbf{I}} - T_t}{T_t} \tag{3}$$

where $m_{\mathbf{I}} = \sum_x \sum_y \mathbf{I}(x, y) / MN$ . The constant $T_t$ is defined as the expected global average brightness for normal natural images. The experimental statistics from several standard image databases show that $T_t$ is appropriate to be set as about the half of maximum pixel intensity, i.e., 128 for 8-bit images. The input image is judged as dimmed if $t < -\tau_t$ , and bright if $t > \tau_t$ , where $\tau_t$ is the threshold used for distinguishing brightness-distorted images from normal ones. The images with normal illuminance $(|t| \le \tau_t)$ are found to be unfit for AGC-based enhancement, thus they would not be addressed by our techniques. Here, $\tau_t$ is set experimentally in consideration of the trade-off between enhancement quality and technical applicability.

In terms of the identified image types of bright and dimmed, the ACG based on negative image and CDF truncation are applied respectively for achieving contrast improvement and brightness restoration.

### *3.1. AGC via Negative Image*

CE of bright images is novelly proposed by applying AGCWD to their negative version. Specifically, the negative image denoted by $\mathbf{I}^{'}$ is formally defined as

$$\mathbf{I}^{'}(x, y) = 255 - \mathbf{I}(x, y) \tag{4}$$

where $x$=1, 2, ..., $M$, $y$=1, 2, ..., $N$. Figs. 4(a)(b) show an example input image and its negative image. We can see from Fig. 4(c) that the pixel intensity distribution of such a negative image is similar as that of a dimmed image. Since large high-value pixels in the primary bright image are reversed by Eq. (4), $\mathbf{I}^{'}$ owns massive low intensity pixels and can be considered as a dimmed image. As a result, the AGCWD can be subsequently applied to $\mathbf{I}'$, which yields an immediate enhanced image $\mathbf{I}_e'$. Lastly, $\mathbf{I}_e'$ is reverted back to positive image space for yielding the final enhanced image $\mathbf{I}_e$ shown in Fig. 4(f).

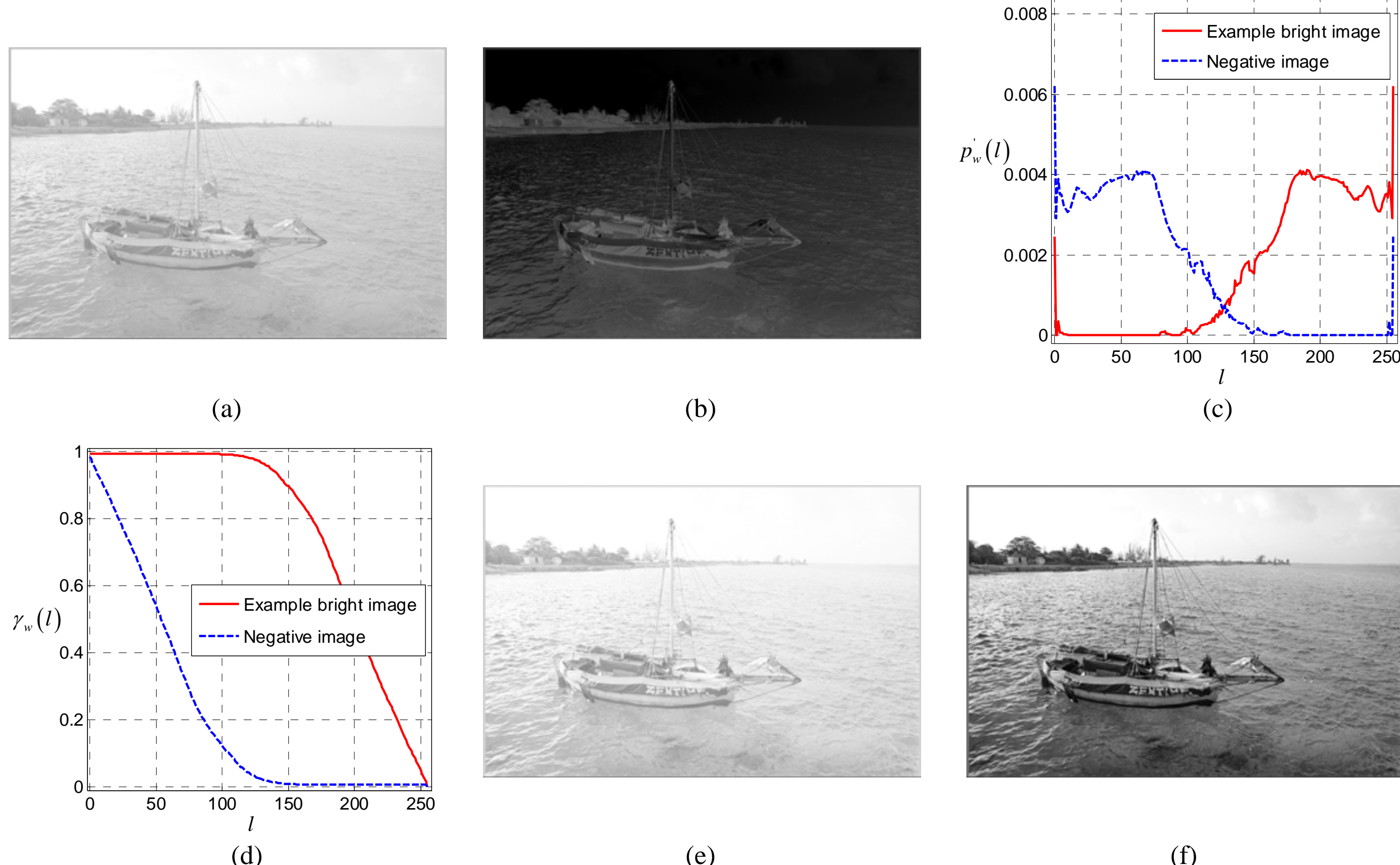

**Fig. 4** Negative-image-based AGC. (a)(b) Example bright image and its negative version; (c) pixel intensity histograms; (d) adaptive gamma curves; (e)(f) enhancement results of ACGWD and negative-image-based AGC, respectively. Here, $\alpha = 0.25$.

We can see that such negative image-based AGC method relies on the successfulness of AGCWD in enhancing the dimmed negative image. In Fig. 4(d), an early and rapid decrease occurs in the adaptive gamma mapping curve generated from the negative image. Consequently, both the brightness and contrast can be enhanced reasonably. However, if AGCWD is directly applied to the bright image, a worse result shown in Fig. 4(e) would be gained due to the used improper mapping function.

Specifically, our proposed negative-image-based AGC algorithm is described briefly in Algorithm-1. It should be mentioned that the weighting factor $\alpha$ used here needs to be addressed elaboratively, since the statistical histogram of natural images may be asymmetric. As such, the setting of $\alpha$ for enhancing bright images is different from that for dimmed images.

| **Algorithm-1: Negative-image-based AGC Algorithm** |
| --- |
| *Step-1*. Obtain the negative image $\mathbf{I}^{'}$ of the input image according to Eq. (4). |
| *Step-2*. Obtain the gray level histogram $p(l)$ of $\mathbf{I}^{'}$, and compute $p_w(l)$ via Eq. (2). |
| *Step-3*. Compute $\gamma_w(l)$=1-$c_w(l)$, where $c_w(l)$ is the CDF derived from normalized $p_w(l)$. |
| *Step-4*. Apply pixel value transformation to $\mathbf{I}^{'}$ according to Eq. (1) and yield $\mathbf{I}_e^{'}$. |
| *Step-5*. Output the enhanced image $\mathbf{I}_e$=$round[255-\mathbf{I}_e^{'}]$, where $round[\cdot]$ is rounding operation. |

### *3.2. AGC via Truncated CDF*

As shown in Fig. 1(b), the input pixels with different intensities are corrected with different gamma values in AGCWD. For example, there are more than 70% pixels within [0, 50] for the primary dimmed image shown in Fig. 1(a). Their corresponding applied gamma decreases rapidly from 1 to around 0.25. As such, the dimmed pixels can be brightened evidently due to the use of relatively small gamma values. That can be seen from the enhanced results shown in Figs. 2(c)(d)(e), which look globally brighter than the original image and the details in dimmed regions become visible. This is the reason why AGCWD works.

However, there exists a serious problem in AGCWD. As shown in Figs. 2(c)(d)(e) (locally magnified versions in Fig. 10), the edges in bright regions, i.e., the white lid of wheels and white shirts, disappear or weaken after CE. Such detail loss phenomenon is typically incurred by AGCWD in enhancing the images with bright regions, which are universal in real applications. This deficiency should attribute to the used overly low gamma values in transforming the median and high intensity pixels. Fig. 5 indicates that $\gamma_w(l)$ continues to monotonously decrease from 0.25 towards 0 in processing the remaining pixels with gray levels within [50, 255]. Especially for the bright regions which typically own high intensity pixels, for example $\mathbf{I}(x, y) > 125$, the applied gamma is below 0.1 so that highly condense pixels move towards 255. Hence, it causes annoying over-saturation and over-enhancement which obscure the edge and structure textures in bright regions.

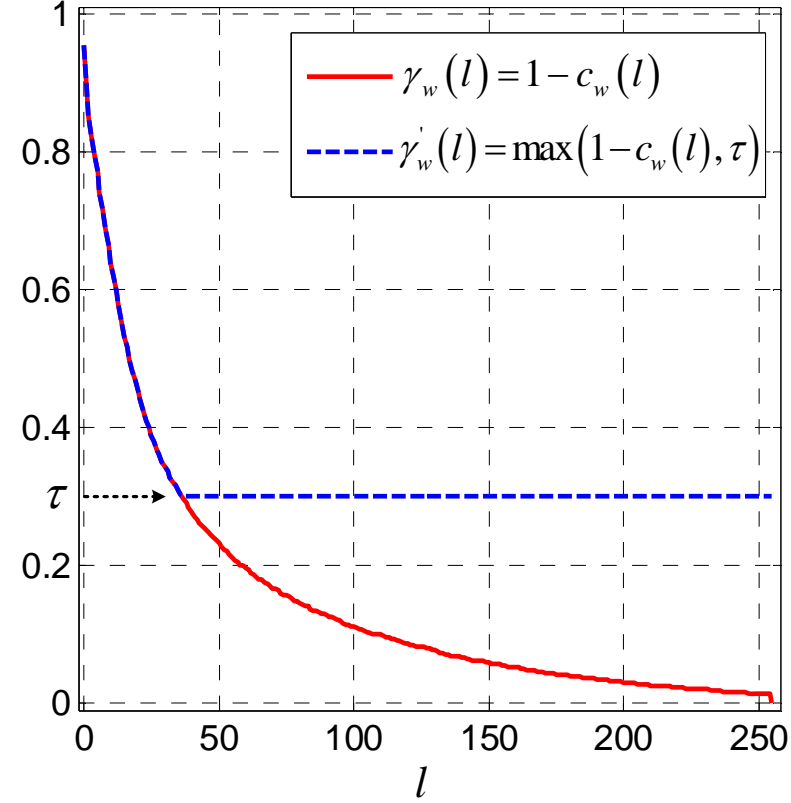

**Fig. 5** Adaptive gamma curves of AGCWD ( $\gamma_w(l)$ ) and CDF-truncated AGC ( $\gamma_w^{'}(l)$ ) for enhancing the image in Fig. 1(a).

The above investigations indicate that the bright regions degraded by AGCWD should attribute to the inappropriate setting of gamma parameter. To attenuate such deficiency, we propose to truncate the CDF curve for limiting it below a reasonable threshold. As such, the corresponding CDF-based gamma can avoid being decreased overly towards zero. Specifically, as illustrated in Fig. 5, we improve the AGCWD method by truncating the adaptive gamma parameter as follows,

$$\gamma_w^{'}(l) = \max\left(\tau,\ 1 - c_w(l)\right) \tag{5}$$

where $\max(\cdot,\cdot)$ is the maximizing operator. $\tau$ is the threshold used for CDF truncation. When $c_w(l)$ is larger than $1-\tau$, $\gamma_w^{'}(l)$ would be boosted to $\tau$. As such, the bright image regions would not be corrected with a rather low gamma value, and the detail loss could be avoided. Through such truncation, $\gamma_w^{'}(l)$ keeps larger than $\tau$ so that the contrast adjustment for bright pixels are restricted reasonably. Meanwhile, the gamma value $\tau$ is also small enough for extending the dynamic range of dimmed pixels.

Overall, our proposed CDF-truncated AGC algorithm is briefly summarized as Algorithm-2. Note that our proposed CE approaches can be extended to enhance color images by applying them to the luminance channel and preserving chrominance channels in a certain color space.

---

**Algorithm-2: CDF-truncated AGC Algorithm**

---

*Step-1*. Obtain gray level histogram $p(l)$ of the input image $\mathbf{I}$.

*Step-2*. Compute $p_w(l)$ according to Eq. (2).

*Step-3*. Compute $\gamma_w^{'}(l)$ according to Eq. (5).

*Step-4*. Output the enhanced image $\mathbf{I}_e$ by transforming $\mathbf{I}$ according to Eq. (1).

---

## 4. Experiments and Discussion

### *4.1. Dataset, Algorithms and Performance Measures*

Both bright and dimmed contrast-distorted images are collected as input test images, which could be simulated or direct camera-outputs. Without loss of generality, the bright and dimmed input images are simulated by respectively applying the gamma corrections with $\gamma = 0.3$ and 2 to the images from Kodak [22], BSD500 [23] and UCID [24] standard databases.

Performance comparison tests are also conducted. The compared CE algorithms include HE [1], HM [4], AGCWD [11], IMADJ (Matlab function '*imadjust*') [10] and SECE [6]. HSV color space is selected and the V channel image is enhanced in all referred tests.

Currently, the performance assessment of CE algorithms is still a challenge task [25-27]. In order to keep consistency with prior works and apply the latest research achievements, the metrics EMEG (Expected Measurement of Enhancement by Gradient) [6], GMSD (Gradient Magnitude Similarity Deviation) [28] and PCQI (Patch-based Contrast Quality Index) [25] are simultaneously used as objective assessment of CE algorithms. EMEG is defined as

$$\text{EMEG}(\mathbf{I}) = \frac{1}{k_1 k_2} \sum_{i=1}^{k_1} \sum_{j=1}^{k_2} \frac{1}{255} \cdot \max\left( \frac{\mathbf{I}_{i,j}^{dx,h}}{\mathbf{I}_{i,j}^{dx,l} + 1}, \frac{\mathbf{I}_{i,j}^{dy,h}}{\mathbf{I}_{i,j}^{dy,l} + 1} \right) \tag{6}$$

where the input image $\mathbf{I}$ is divided into $k_1k_2$ overlapping sub-blocks $\mathbf{I}_{i,j}$ of size $8 \times 8$ pixels. $\mathbf{I}_{i,j}^{dx,h}$, $\mathbf{I}_{i,j}^{dx,l}$ are respectively the highest and lowest values of absolute row-based block derivative, while $\mathbf{I}_{i,j}^{dy,h}$, $\mathbf{I}_{i,j}^{dy,l}$ are the column-based ones. The blocks with high contrast own high EMEG values, where the maximum value is 1. Contrarily, EMEG of smooth blocks is relatively low and reaches as low as 0. That is, $\text{EMEG}(\mathbf{I}) \in [0,1]$. Although EMEG is sensitive to noise, it is still expected that $\text{EMEG}(\mathbf{I}_e) > \text{EMEG}(\mathbf{I})$ [6].

GMSD is a full-reference image quality assessment for measuring the perceptual similarity between two images via gradient comparison [28]. Here, $\text{GMSD}(\mathbf{I}_e, \mathbf{I}_r)$ denotes the similarity between the enhanced image $\mathbf{I}_e$ and the reference image $\mathbf{I}_r$. The GMSD value is lower if the difference is less, and becomes 0 if corresponding gradient images are the same. However, we discover that GMSD may fail to reflect the discrepancy of mean intensities between two images, which determine the global perceptual brightness.

PCQI provides an accurate prediction on the human perception of contrast variations between the enhanced and reference images [25]. The PCQI is defined as

$$\text{PCQI}(\mathbf{I}_e, \mathbf{I}_r) = P_c \cdot P_s \cdot P_i \tag{7}$$

where $P_c \in [0, 2]$ measures the contrast change of $\mathbf{I}_e$ comparing with $\mathbf{I}_r$. Contrast increases if $P_c > 1$, and the higher $P_c$ value implies more improvement. $P_s \in [0, 1]$ and $P_i \in [0, 1]$ respectively measure the distortions of image structure and mean intensity between $\mathbf{I}_e$ and $\mathbf{I}_r$. The higher $P_s$, $P_i$ values mean less distortions. The PCQI ($P$) value provides the overall evaluation metric.

Based on plentiful experimental observations and quantitative verifications, the involved parameters in our proposed algorithm are experimentally set as $T_t = 112$, $\tau_t = 0.3$, $\tau = 0.5$, $\alpha = 0.25$ for enhancing bright images and $\alpha = 0.75$ for dimmed ones. Such setting could be easily adjusted within a limited range without apparently affecting the enhancement results.

*4.2. Qualitative Assessments on Bright Images*

Qualitative assessment results on the simulated bright type of low-contrast images are shown in Figs. 6~9 and Tables 1, 2.

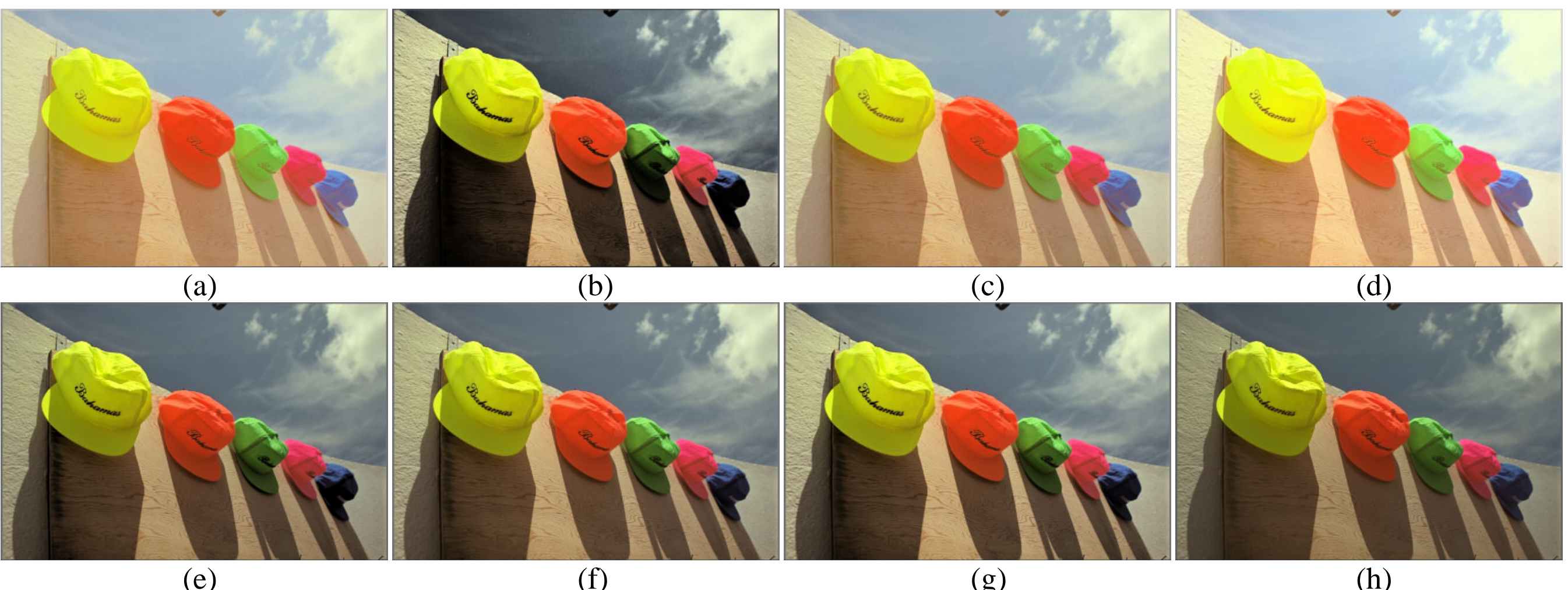

**Fig. 6** Results for **bright** image *Hats*. (a) Input bright image; Enhanced images obtained by (b) HE, (c) HM, (d) AGCWD, (e) IMADJ, (f) SECE, (g) Our proposed algorithms; (h) Original version in Kodak dataset.

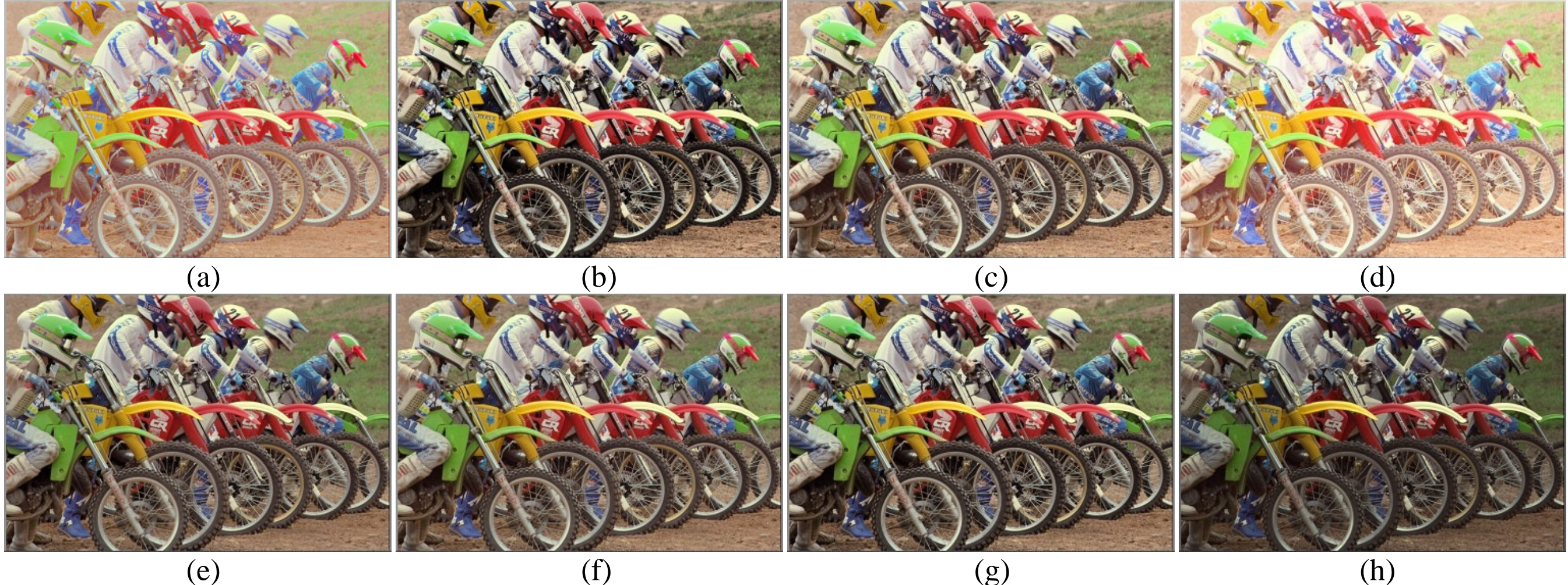

**Fig. 7** Results for **bright** image *Moto*. (a) Input bright image; Enhanced images obtained by (b) HE, (c) HM, (d) AGCWD, (e) IMADJ, (f) SECE, (g) Our proposed algorithms; (h) Original version in Kodak dataset.

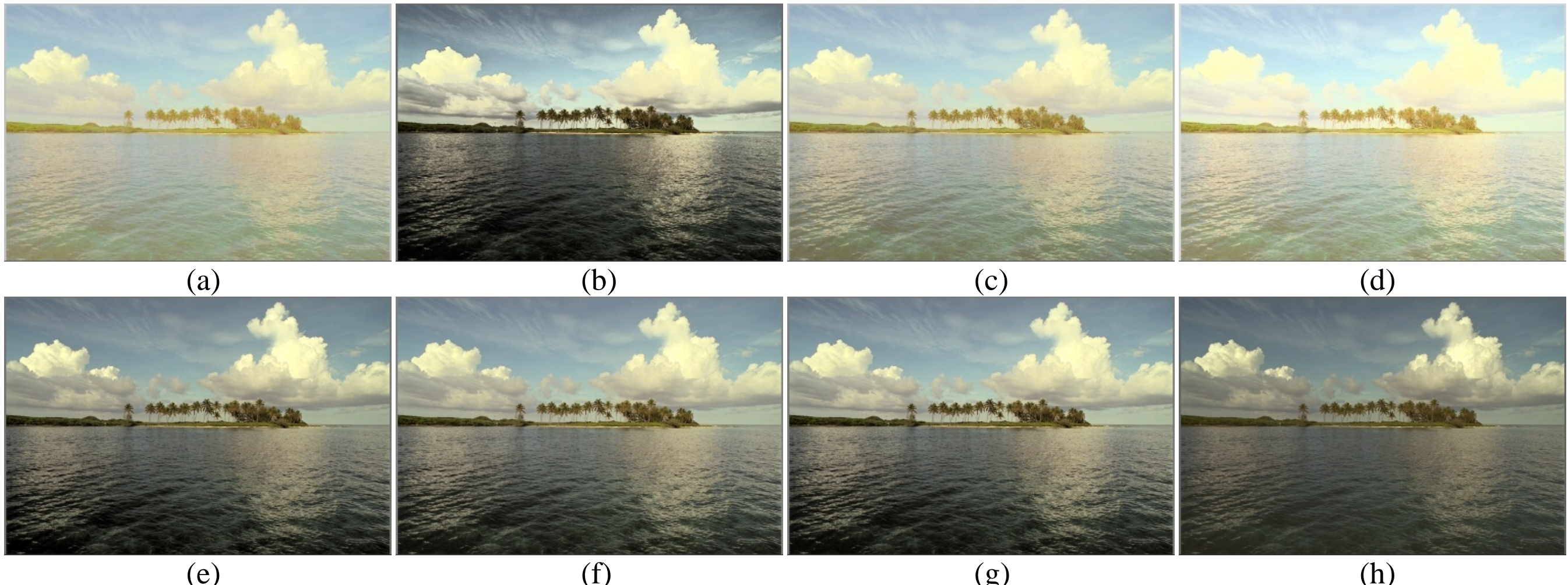

**Fig. 8** Results for **bright** image *Island*. (a) Input bright image; Enhanced images obtained by (b) HE, (c) HM, (d) AGCWD, (e) IMADJ, (f) SECE, (g) Our proposed algorithms; (h) Original version in Kodak dataset.

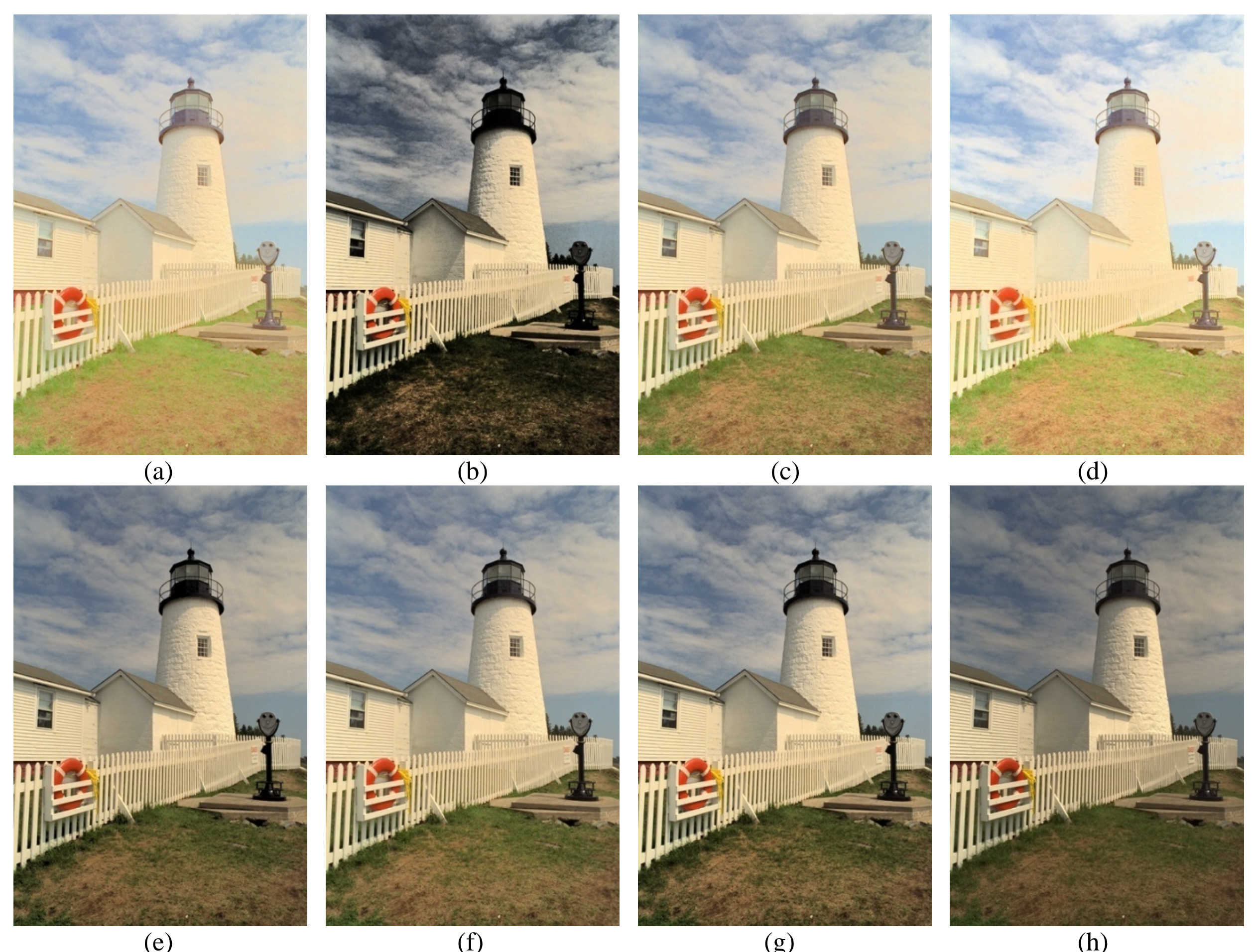

**Fig. 9** Results for **bright** image *Tower*. (a) Input bright image; Enhanced images obtained by (b) HE, (c) HM, (d) AGCWD, (e) IMADJ, (f) SECE, (g) Our proposed algorithms; (h) Original version in Kodak dataset.

**Table 1** EMEG ($E$) and GMSD ($G$) values ($\times 10^{-3}$) of the **bright** sample images (Figs. 6~9) enhanced by different algorithms.

| Algo. | Input | | HE | | HM | | AGCWD | | IMADJ | | SECE | | Prop. | |
|---|---|---|---|---|---|---|---|---|---|---|---|---|---|---|
| Metric | $E$ | $G$ | $E$ | $G$ | $E$ | $G$ | $E$ | $G$ | $E$ | $G$ | $E$ | $G$ | $E$ | $G$ |
| *Hats* | 43 | 84 | 133 | 99 | 54 | 76 | 52 | 150 | 89 | 31 | 88 | 23 | 99 | 43 |
| *Moto* | 152 | 87 | 378 | 92 | 309 | 83 | 199 | 182 | 262 | 51 | 262 | 39 | 301 | 44 |
| *Island* | 70 | 81 | 192 | 98 | 101 | 79 | 93 | 146 | 157 | 70 | 155 | 62 | 164 | 62 |
| *Tower* | 75 | 98 | 224 | 119 | 120 | 73 | 89 | 200 | 158 | 30 | 147 | 28 | 169 | 39 |

**Table 2** PCQI ($P$) values ($\times 10^{-3}$) of the **bright** sample images (Figs. 6~9) enhanced by different algorithms.

| Algo. | Input | | | | HE | | | | HM | | | | AGCWD | | | | IMADJ | | | | SECE | | | | Prop. | | | |
|---|---|---|---|---|---|---|---|---|---|---|---|---|---|---|---|---|---|---|---|---|---|---|---|---|---|---|---|---|
| Metric | $P_c$ | $P_s$ | $P_i$ | $P$ | $P_c$ | $P_s$ | $P_i$ | $P$ | $P_c$ | $P_s$ | $P_i$ | $P$ | $P_c$ | $P_s$ | $P_i$ | $P$ | $P_c$ | $P_s$ | $P_i$ | $P$ | $P_c$ | $P_s$ | $P_i$ | $P$ | $P_c$ | $P_s$ | $P_i$ | $P$ |
| *Hats* | 752 | 992 | 732 | 538 | 1210 | 974 | 885 | 1048 | 841 | 988 | 750 | 617 | 800 | 975 | 674 | 515 | 1080 | 986 | 952 | 1015 | 1076 | 988 | 927 | 987 | 1124 | 986 | 915 | 1017 |
| *Moto* | 750 | 992 | 709 | 520 | 1259 | 979 | 876 | 1089 | 1159 | 983 | 806 | 923 | 846 | 961 | 642 | 517 | 1084 | 992 | 880 | 948 | 1090 | 993 | 868 | 941 | 1171 | 993 | 899 | 1048 |
| *Island* | 754 | 993 | 721 | 534 | 1245 | 982 | 888 | 1093 | 909 | 989 | 732 | 653 | 848 | 980 | 659 | 539 | 1165 | 988 | 932 | 1078 | 1148 | 990 | 903 | 1033 | 1184 | 990 | 910 | 1073 |
| *Tower* | 687 | 993 | 744 | 500 | 1226 | 972 | 904 | 1083 | 917 | 987 | 791 | 709 | 754 | 970 | 691 | 491 | 1056 | 990 | 933 | 977 | 1025 | 991 | 890 | 904 | 1110 | 990 | 912 | 1004 |

Fig. 6 shows the results on the image *Hats*, where (a) indicates the bright image to be enhanced, and (b)~(g) indicate the enhanced results generated by different CE algorithms. As shown in Fig. 6(b), although HE achieves high contrast, it incurs over-enhancement in shadow regions which look much black. Figs. 6(c)(d) show that the contrast improvements yielded by HM and AGCWD are rather weak and even void. IMADJ, SECE and our proposed CE algorithm have successfully boost the contrast of distorted images without incurring annoying artifacts, and the enhanced results are visually consistent with the unaltered standard image shown in Fig. 6(h).

Such visual results are consistently validated by corresponding objective performance measurements given in Tables 1, 2. Comparing with AGCWD and HM, our method attains higher EMEG, $P_c$ values, lower GMSD and higher $P_s$, $P_i$ values. Such digitals reveal that more contrast improvement and less structure/intensity distortions are achieved by our method. Although HE gains high EMEG, $P_c$ values, it has high GMSD and low $P_s$, $P_i$ values which imply large distortions. In terms of performance tradeoff, ours is comparative with IMADJ and SECE, since approximate measurements have been obtained by such methods. Note that slightly higher contrast improvement may be at the cost of a little more distortions, i.e., either the structure or the mean intensity distortion.

Similar results are also obtained on the other example images *Moto*, *Island* and *Tower*, as shown in Figs. 7, 8 and 9, respectively. We can see that quality of the images enhanced by our method always keeps comparative with the state of the art, i.e., SECE, and the classical IMADJ method. HE and HM could not work well on all sample images, while AGCWD behaves the most badly due to the least contrast increase. Such consistent conclusions could also be verified by the corresponding results shown in Tables 1, 2.

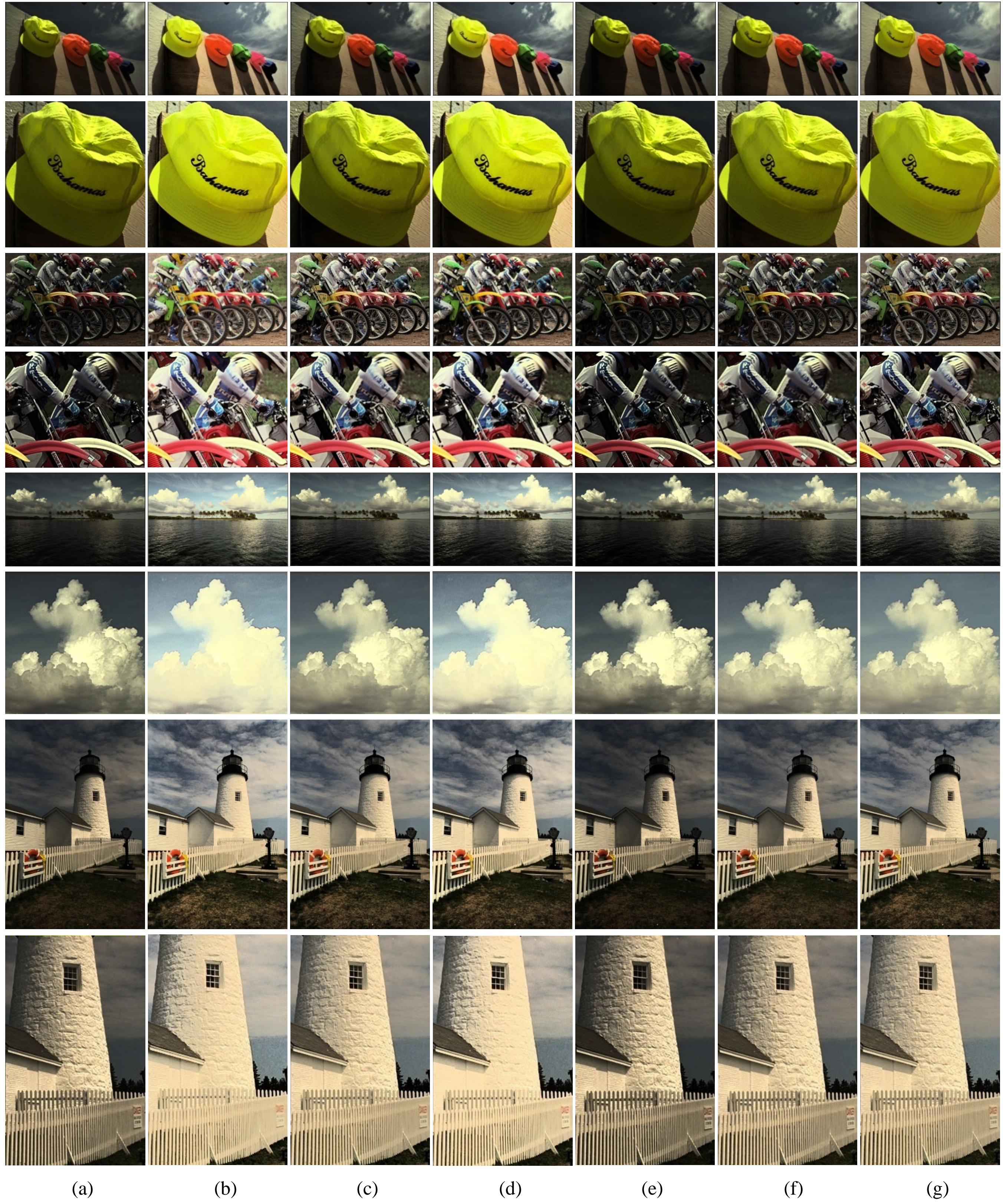

(a) (b) (c) (d) (e) (f) (g)

**Fig. 10** Results for **dimmed** images *Hats*, *Moto*, *Island* and *Tower*. (a) Input dimmed image; Enhanced images obtained by (b) HE, (c) HM, (d) AGCWD, (e) IMADJ, (f) SECE, (g) Our proposed algorithms. The corresponding locally magnified versions are shown in even rows.

**Table 3** EMEG ($E$) and GMSD ($G$) values ($\times 10^{-3}$) of the **dimmed** sample images (Fig. 10) enhanced by different algorithms.

| Algo. | Input | | HE | | HM | | AGCWD | | IMADJ | | SECE | | Prop. | |
|---|---|---|---|---|---|---|---|---|---|---|---|---|---|---|
| Metric | $E$ | $G$ | $E$ | $G$ | $E$ | $G$ | $E$ | $G$ | $E$ | $G$ | $E$ | $G$ | $E$ | $G$ |
| *Hats* | 80 | 55 | 130 | 97 | 95 | 56 | 129 | 101 | 82 | 64 | 96 | 45 | 101 | 50 |
| *Moto* | 208 | 86 | 375 | 91 | 354 | 69 | 393 | 104 | 209 | 87 | 252 | 43 | 294 | 53 |
| *Island* | 89 | 63 | 189 | 97 | 123 | 50 | 184 | 108 | 102 | 67 | 127 | 43 | 147 | 59 |
| *Tower* | 146 | 56 | 220 | 117 | 186 | 59 | 215 | 92 | 149 | 62 | 161 | 31 | 183 | 71 |

**Table 4** PCQI ($P$) values ($\times 10^{-3}$) of the **dimmed** sample images (Fig. 10) enhanced by different algorithms.

| Algo. | Input | | | | HE | | | | HM | | | | AGCWD | | | | IMADJ | | | | SECE | | | | Prop. | | | |
|---|---|---|---|---|---|---|---|---|---|---|---|---|---|---|---|---|---|---|---|---|---|---|---|---|---|---|---|---|
| Metric | $P_c$ | $P_s$ | $P_i$ | $P$ | $P_c$ | $P_s$ | $P_i$ | $P$ | $P_c$ | $P_s$ | $P_i$ | $P$ | $P_c$ | $P_s$ | $P_i$ | $P$ | $P_c$ | $P_s$ | $P_i$ | $P$ | $P_c$ | $P_s$ | $P_i$ | $P$ | $P_c$ | $P_s$ | $P_i$ | $P$ |
| *Hats* | 966 | 995 | 804 | 774 | 1208 | 979 | 887 | 1054 | 1061 | 992 | 828 | 875 | 1190 | 985 | 893 | 1053 | 976 | 993 | 790 | 767 | 1074 | 992 | 859 | 919 | 1110 | 989 | 939 | 1030 |
| *Moto* | 814 | 987 | 824 | 659 | 1253 | 978 | 876 | 1083 | 1222 | 982 | 932 | 1124 | 1232 | 974 | 915 | 1106 | 814 | 987 | 821 | 658 | 983 | 989 | 880 | 857 | 1121 | 984 | 947 | 1045 |
| *Island* | 882 | 995 | 808 | 708 | 1242 | 989 | 890 | 1100 | 1058 | 994 | 872 | 919 | 1201 | 987 | 893 | 1067 | 947 | 992 | 813 | 765 | 1073 | 994 | 893 | 955 | 1103 | 992 | 943 | 1031 |
| *Tower* | 962 | 995 | 810 | 778 | 1224 | 978 | 905 | 1090 | 1189 | 991 | 933 | 1099 | 1213 | 982 | 909 | 1090 | 969 | 994 | 801 | 775 | 1083 | 995 | 878 | 948 | 1160 | 988 | 941 | 1078 |

### *4.3. Qualitative Assessments on Dimmed Images*

In this subsection, the corresponding qualitative assessment results on dimmed low-contrast images are illustrated in Fig. 10 and Tables 3, 4.

We can see from the locally magnified regions that both HE and AGCWD methods typically incur over-enhancement on bright regions, such as top of the hats, edges within the white circle lid and shirts, contours of the clouds and textures of the chimney. Details in such image regions are lost to some extent after CE. The corresponding distortion measurements $P_s$, $P_i$ are relatively low. For example on the image *Island*, $P_s$=0.989, $P_i$=0.890 for HE and $P_s$=0.987, $P_i$=0.893 for AGCWD. On the contrary, our proposed CE method avoids such over-enhancement and has $P_s$=0.992, $P_i$=0.943, which are more close to 1 and signify less distortions. Although such better visual quality is achieved, the given overall metric $P$ for our method is instead lower than those of HE and AGCWD due to the affection of $P_c$. As a result, we believe that the PCQI sub-metrics are more reasonable to be treated separately.

A good CE method should achieve a balance between contrast improvement and image distortion. Less distortion should be incurred while more contrast improvement is gained. In terms of such viewpoint, our proposed method could achieve the comparative results with SECE and HM. Note that HM causes slightly apparent global intensity distortion on *Hats* and *Island*, which are validated by the low $P_i$ values, 0.828 and 0.872, respectively. Although IMADJ could boost image contrast adequately, it causes more obvious global intensity distortion and the enhanced images are still rather black.

**Table 5** Average EMEG ($E$) and GMSD ($G$) values ($\times 10^{-3}$) of the test images from BSD500 enhanced by different algorithms.

| Algo. | Input | | HE | | HM | | AGCWD | | IMADJ | | SECE | | Prop. | |
|---|---|---|---|---|---|---|---|---|---|---|---|---|---|---|
| Metric | $E$ | $G$ | $E$ | $G$ | $E$ | $G$ | $E$ | $G$ | $E$ | $G$ | $E$ | $G$ | $E$ | $G$ |
| Bright | 113 | 53 | 265 | 114 | 170 | 49 | 135 | 127 | 188 | 44 | 187 | 30 | 205 | 42 |
| Dimmed | 162 | 107 | 270 | 123 | 211 | 100 | 270 | 111 | 193 | 108 | 202 | 77 | 208 | 83 |

**Table 6** Average PCQI ($P$) values ($\times 10^{-3}$) of the test images from BSD500 enhanced by different algorithms.

| Algo. | Input | | | | HE | | | | HM | | | | AGCWD | | | | IMADJ | | | | SECE | | | | Prop. | | | |
|---|---|---|---|---|---|---|---|---|---|---|---|---|---|---|---|---|---|---|---|---|---|---|---|---|---|---|---|---|
| Metric | $P_c$ | $P_s$ | $P_i$ | $P$ | $P_c$ | $P_s$ | $P_i$ | $P$ | $P_c$ | $P_s$ | $P_i$ | $P$ | $P_c$ | $P_s$ | $P_i$ | $P$ | $P_c$ | $P_s$ | $P_i$ | $P$ | $P_c$ | $P_s$ | $P_i$ | $P$ | $P_c$ | $P_s$ | $P_i$ | $P$ |
| Bright | 825 | 995 | 835 | 689 | 1232 | 965 | 892 | 1063 | 1014 | 989 | 866 | 871 | 868 | 976 | 763 | 643 | 1094 | 993 | 918 | 997 | 1096 | 994 | 932 | 1015 | 1126 | 992 | 923 | 1033 |
| Dimmed | 863 | 984 | 815 | 696 | 1227 | 957 | 880 | 1040 | 1065 | 977 | 872 | 919 | 1183 | 972 | 891 | 1034 | 965 | 982 | 834 | 800 | 1042 | 984 | 878 | 908 | 1087 | 978 | 902 | 964 |

### *4.4. Quantitative Assessments*

In order to quantitatively evaluate the algorithm performance, bright and dimmed contrast-distorted test images are prepared by respectively applying gamma correction with $\gamma = 0.3$ and 2 to the 500 images from BSD500 dataset. Tables 5, 6 show the average metrics for evaluating the quality of images enhanced by different CE methods. Through observing average EMEG and $P_c$, we can see that all CE methods could improve the contrast of input images, and our method is even better than SECE and IMADJ. Nevertheless, the average GMSD and $P_s$ of our method are slightly worse than those of SECE and IMADJ. That is to say, our method could boost the contrast more remarkably at the cost of weakly intenser image distortion. Besides, we could note that $P_i$ of our method on dimmed images (0.902) is higher than those of SECE (0.878) and IMADJ (0.834). Overall, such results demonstrate that the performance of our proposed CE method is comparative with the state-of-the-art CE methods.

Simultaneously, remarkable improvement between AGCWD and our method could be seen from such statistical results. As for bright images, the $P_c$ value increases from 0.868 (AGCWD) to 1.126 (Prop.), and $P_s$, $P_i$ increase from 0.976, 0.763 to 0.992, 0.923, respectively. As for the dimmed images, although AGCWD gets higher $P_c$, it also incurs more serious distortions which can be seen from the relatively small $P_s$, $P_i$ values. EMEG and GMSD metrics shown in Table 5 also consistently validate such conclusions.

The same quantitative tests are also conducted on Kodak and UCID datasets, and consistent results have been obtained. Our method is also testified to be effective.

*4.5. Computation Time*

A good CE algorithm also requires low computational complexity. We also evaluate the algorithm complexity of our proposed method. All algorithms are run on a computer with Intel Core i5-5200U CPU @2.2 GHz and 8G RAM. The applied software platform is MATLAB R2013a. The average computation time used for enhancing a test image created from BSD500 is computed.

As displayed in Table 7, the resulting average computation time per image of our proposed method is 28.6 ms, which is comparative with that of AGCWD. The additional operations in our proposed method, i.e., negative image transformation and CDF truncation, do not incur rapid increasement of computational complexity. SECE has a higher time complexity (38.6 ms) than ours. The histogram based approaches including HE and HM are testified to own the merit of low complexity. Note that IMADJ is the most fast with 1.8 ms per image, which is far below other methods. Such a result attributes to the simplicity of the involved data operations, which just comprise simple statistic and stretching of pixel intensities. Moreover, the prior results show that impressive visual enhancement effects can be achieved by IMADJ. Generally, all CE techniques pursue the same goal of achieving more contrast increasement with less image distortion at the cost of less computational resources.

**Table 7** Average computation times (ms) of different CE algorithms per test image created from BSD500.

| Algo. | HE | HM | AGCWD | IMADJ | SECE | Prop. |
| --- | --- | --- | --- | --- | --- | --- |
| Time | 12.4 | 11.8 | 26.4 | 1.8 | 38.6 | 28.6 |

*4.6. CE on Real Distorted Images*

We also tested the CE performance on real contrast-distorted images, which are selected from UCID dataset in terms of the abnormal brightness. Fig. 11 illustrates the results on the bright type of distorted images. The high global brightness of such input images is expected to be decreased by CE operations. The results demonstrate the effectiveness of our proposed method, which outperforms AGCWD and HE distinctly, and resembles the other approaches. Specifically, we could pay attention to the white regions within five example images, i.e., papers with characters, the skirt of a bear doll, cloth in background, text on book sides and flowers on the ground, respectively. For example, the wrinkle textures in the skirt and cloth regions are eliminated to some extent by AGCWD, but yet preserved well by our method.

The enhanced results on real dimmed images are shown in Fig. 12. We can see that HE often incurs globally over-enhancement and some annoying artifacts in local regions, such as the sky area. As the prior tests on simulated images, AGCWD still behaves badly on white and bright regions, for an instance, the tower in the first example image and the billboard with characters 'mm' in the second one. Comparing with the results of ours, more textured structure details of tower tops are lost, and the characters are more blurry. Note that IMADJ and SECE are both inclined to bring under-enhancement, which could not heighten the global brightness properly and the enhanced images are still rather dimmed. Nevertheless, our method and HM could efficiently avoid such defects, and in general, superior enhancement effects are achieved.

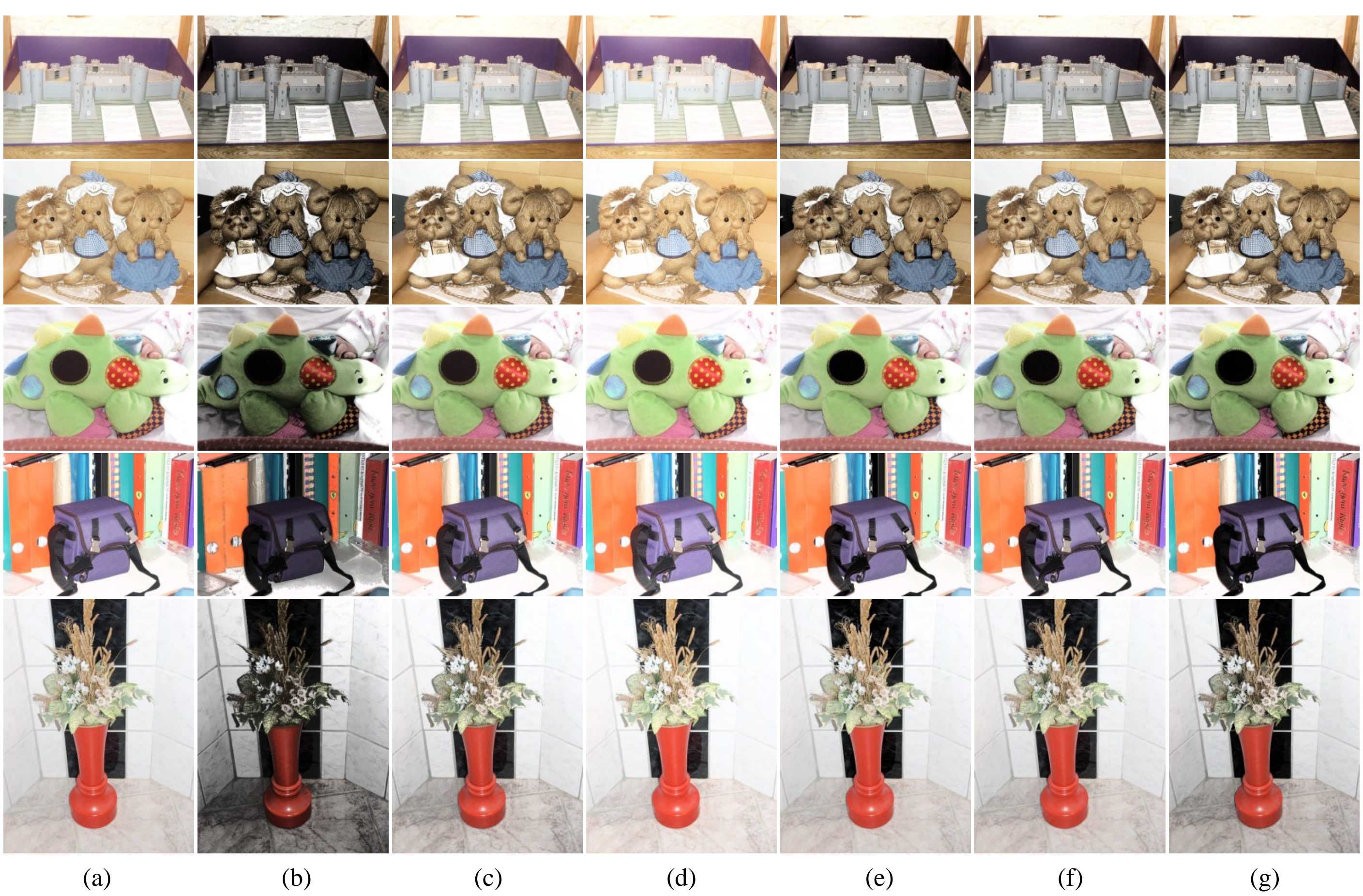

**Fig. 11** Results for real **bright** images. (a) Input bright image; Enhanced images obtained by (b) HE, (c) HM, (d) AGCWD, (e) IMADJ, (f) SECE, (g) Our proposed algorithms.

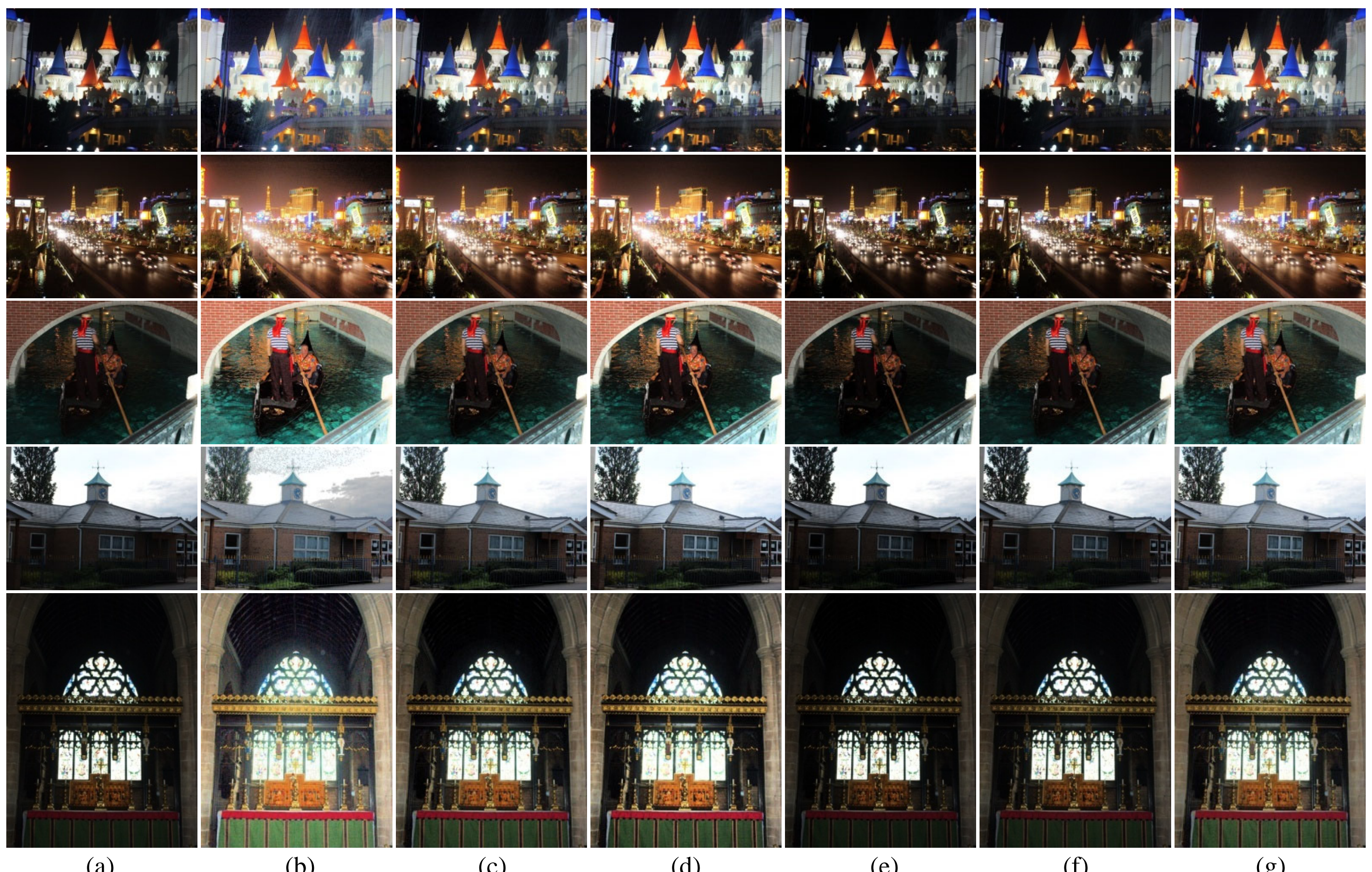

**Fig. 12** Results for real **dimmed** images. (a) Input dimmed image; Enhanced images obtained by (b) HE, (c) HM, (d) AGCWD, (e) IMADJ, (f) SECE, (g) Our proposed algorithms.

## 5. Conclusion

A new effective and efficient image contrast enhancement method is proposed based on an improved adaptive gamma correction. The methodology of negative images is used to enhance the contrast of bright images. CDF truncation is proposed to reconstruct the intensity-sensitive adaptive gamma for improving the enhancement effects on dimmed images. Extensive qualitative and quantitative experiments show that our proposed scheme achieves better or comparative enhancement effects than previous techniques. The contrast of both bright and dimmed input images is enhanced effectively and efficiently without incurring annoying artifacts. In the future work, we would try to improve the capability of our proposed method in enhancing more types of images, instead of limiting to dimmed and bright ones.